\documentclass{pasa}%

\usepackage{graphicx}
\usepackage{url}

\title[Galaxy spin direction asymmetry in JWST]{Galaxy spin direction asymmetry in JWST deep fields}

\author[Lior Shamir]{Lior Shamir
\affil{Kansas State University \\ Manhattan, KS 66506}
}

\jid{PASA}
\doi{10.1017/pasa.2024.20}
\jyear{\the\year}
\usepackage{aas_macros}
\usepackage[pdftex,colorlinks=true,linkcolor=darkblue,citecolor=darkred,urlcolor=blue]{hyperref}
\hypersetup{colorlinks,citecolor=blue,linkcolor=blue,urlcolor=blue}

\begin{document}

\begin{frontmatter}

\maketitle

\begin{abstract}

The unprecedented imaging power of JWST provides new abilities to observe the shapes of objects in the early Universe in a way that has not been possible before. 
Recently, JWST acquired a deep field image inside the same field imaged in the past as the HST Ultra Deep Field. Computer-based quantitative analysis of spiral galaxies in that field shows that among 34 galaxies for which their rotation of direction can be determined by the shapes of the arms, 24 rotate clockwise, and just 10 rotate counterclockwise. The one-tailed binomial distribution probability to have asymmetry equal or stronger than the observed asymmetry by chance is $\sim$0.012. While the analysis is limited by the small size of the data, the observed asymmetry is aligned with all relevant previous large-scale analyses from all premier digital sky surveys, all show a higher number of galaxies rotating clockwise in that part of the sky, and the magnitude of the asymmetry increases as the redshift gets higher. This paper also provides data and analysis to reproduce previous experiments suggesting that the distribution of galaxy rotation in the Universe is random, to show that the exact same data used in these studies in fact show non-random distribution, and in excellent agreement with the results shown here. These findings reinforce consideration of the possibility that the directions of rotation of spiral galaxies as observed from Earth are not necessarily randomly distributed. The explanation can be related to the large-scale structure of the Universe, but can also be related to a possible anomaly in the physics of galaxy rotation.



\end{abstract}

\begin{keywords}
Galaxies: spiral -- Cosmology: large-scale structure of universe -- Cosmology: Cosmic anisotropy
\end{keywords}

\end{frontmatter}


\section{Introduction}
\label{introduction}

James Webb Space Telescope (JWST) is the most powerful imaging device in the history of astronomy, with unprecedented ability to image astronomical objects in the early Universe. The ability to observe the deep Universe in fine detail can provide new insights about the origin of galaxies and the nature of the early Universe.

According to current theories, the number of galaxies rotating in opposite directions is expected to be the same within statistical fluctuations in any given field in the sky. That is, the number of galaxies rotating in one direction is expected to be equal within statistical error to the number of galaxies rotating in the opposite direction. That assumption, however, has been challenged by observations of space-based \citep{shamir2020pasa,shamir2021aas} and Earth-based \citep{macgillivray1985anisotropy,longo2011detection,shamir2012handedness,shamir2016asymmetry,shamir2019large,shamir2020patterns,shamir2020large,shamir2020pasa,shamir2021particles,shamir2021large,shamir2022new,shamir2022large,shamir2022asymmetry,shamir2022analysis} instruments. These studies were based on a large number of galaxies and several different instruments, showing that the number of galaxies spinning in opposite directions is not necessarily the same in all fields. Sky surveys with large footprints also showed a higher number of galaxies spinning clockwise in one hemisphere, while in the opposite hemisphere the number of galaxies spinning counterclockwise was higher, forming a cosmological-scale dipole axis \citep{shamir2021large,shamir2022new,shamir2022asymmetry,shamir2022analysis}.  

On the other hand, several studies also suggested that the number of galaxies rotating in opposite directions is distributed randomly \citep{iye1991catalog,land2008galaxy,hayes2017nature,tadaki2020spin,iye2020spin}. Analysis and reproduction of all of these experiments showed that the data on which these reports are based is in fact in agreement with non-random distribution \citep{sym15091704}. The analysis and reproduction, including the code and data to reproduce the experiments, are described in detail in \cite{sym15091704,psac058,mcadam2023reanalysis,shamir2022analysis}, and will be explained briefly also in Section~\ref{other_reports} of this paper.

While several analyses using different instruments were performed, the observation has not been studied in the early Universe as imaged by JWST. Since the magnitude of the asymmetry has been identified to grow as the redshift gets larger \citep{shamir2019large,shamir2020patterns,shamir2022large}, studying the asymmetry in deep fields can lead to new observations. This paper examines the possibility of an anomaly in the distribution of galaxies rotating in opposite directions in the JWST deep field taken inside the Hubble Space Telescope (HST) Ultra Deep Field. The observation is compared to analyses with other space-based and Earth-based telescopes that image the same field, as well as to analyses of the entire sky. 



\section{Observed asymmetry in spiral galaxies observed through JWST}
\label{results}

The unprecedented sensitivity of JWST provides sufficient detail to identify the spin directions of early galaxies. The direction of rotation of these galaxies can be determined by the shape of the arms. That can provide the projected angular momentum of the stellar mass as seen from Earth. Although some galaxies can have leading arms \citep{byrd2019ngc}, such galaxies are very rare, and in the vast majority of the cases spiral galaxies have trailing arms. The shape of the arms therefore allows the identification of the direction of rotation of spiral galaxies as observed from Earth.

The primary data used in this study is the JWST deep field image taken inside the field of the Hubble Ultra Deep Field. The centre of the field is at around ${(\alpha=53.2^o, \delta=-28^o)}$. The image and information that describes it is publicly available\footnote{ \url{https://webbtelescope.org/contents/media/images/01GXE4A07MB2RG6GHDGF3CHHJ4/}}. The image is also displayed in Figure~\ref{jwst} below. The file is a TIF file that combines the F182M, F210M, F430M, F460M, and F480M filters. The image was taken in October 2022, and released to the public on April 2023.

The shape of the arms is in many cases complex and unclear, and many in-between cases can bias the results. Therefore, manual annotation of galaxies can be biased by the human perception. For that reason, a fully symmetric computer analysis of the spin direction was used. The results can be inspected by the human eye, but it is important that the annotation is performed with no manual intervention, as any such intervention might lead to bias.

The automatic analysis was done by using the Ganalyzer algorithm \citep{shamir2011ganalyzer,ganalyzer_ascl}. In summary, the Ganalyzer algorithm first applies basic object detection and separates all objects that are larger than 400 foreground pixels. Then, each such extended object is separated from the image and converted into its radial intensity plot. 

The radial intensity plot of each object is a 35$\times$360 matrix, where the intensity of the pixel at Cartesian coordinates $(x,y)$ is the median intensity of the 5$\times$5 pixels around Cartesian coordinates $(O_x+\sin(\theta) \cdot r,O_y-\cos(\theta)\cdot r)$ in the original image, where {\it r} is the radial distance, $(O_x,O_y)$ are the coordinates of the galaxy centre, and $\theta$ is the polar angle. The radius is the percentage of the galaxy radius, and the polar angle is measured in degrees \citep{shamir2011ganalyzer}.

Because the arms are brighter than the background, arm pixels are brighter than pixels that are not part of the arm. Therefore, the arm pixels can be detected by applying a 2D peak detection algorithm \citep{morhavc2000identification} to each line in the radial intensity plot. A linear regression is applied to the peaks in neighbouring lines, and the sign of the slope of the linear regression is used to determine the direction of the arm curves. Consequently, the directions of the curves is used to determine the direction of rotation of each galaxy. The direction of rotation is determined only if the galaxy has at least 30 peaks detected, otherwise the galaxy is determined as an elliptical galaxy or another form that does not have clear identifiable direction of rotation. To avoid cases in which the curve is too mild to determine the direction of rotation, the slope of the linear regression needs to be greater than 0.35, otherwise the arm is ignored \citep{shamir2011ganalyzer}. The analysis process and experimental results are also described in \citep{shamir2011ganalyzer,shamir2013color,shamir2016asymmetry,shamir2017photometric,shamir2017large,shamir2017colour,shamir2020large,shamir2020pasa,shamir2021particles,shamir2021large,shamir2022new,shamir2022large,shamir2022analysis}.  

Ganalyzer is a deterministic algorithm that is based on defined symmetric rules. It does not use machine learning, deep learning, or any other method that is based on complex rules determined automatically from data. The simple ``mechanical'' nature of Ganalyzer ensures that it is not subjected to biases of machine learning systems that are often very difficult to identify, and are common in machine learning systems \citep{DHAR202192,ball2023}, and specifically in astronomy \citep{DHAR2022100545}. 

{\it Ganalyzer} can be applied without the need to first select spiral galaxies. That is because {\it Ganalyzer} inspects the arms, and if no arms are found, or if the arms are not curved, the galaxy is not annotated and therefore not used in the analysis. This is different from some machine learning methods that require a first step of selection of spiral galaxies before the spin directions can be analysed. That selection might not be fully symmetric when done by machine learning methods, and in any case the symmetry of these methods, especially when using deep neural networks, is difficult to verify \citep{DHAR202192,ball2023}.
 
Quantitative analysis of the accuracy of the {\it Ganalyzer} algorithm can be found in previous experiments \citep{shamir2020asymmetry,shamir2022asymmetry}. In both cases 200 galaxies that were annotated as spinning clockwise and 200 galaxies annotated as spinning counterclockwise were selected randomly and examined manually. In both cases no galaxy that was annotated as spinning clockwise was observed manually as spinning counterclockwise, or vice versa. Perhaps the large-scale dataset that is the most similar to JWST is the analysis of HST COSMOS galaxies \citep{shamir2020asymmetry}. The manual examination of the 400 randomly selected galaxies did not identify an incorrectly annotated galaxy, and not even a case where the classification was not sufficiently clear. In all cases the annotations of the galaxies were clear and correct. The dataset of annotated COSMOS galaxies is publicly available\footnote{\url{http://people.cs.ksu.edu/~lshamir/data/assym_COSMOS}}. The high accuracy of the annotation comes at the cost of completeness. That is, the annotations are accurate, but most of the galaxies in the original dataset are not assigned with a direction of rotation, and are therefore not used in the final analysis \citep{shamir2013color,shamir2016asymmetry,shamir2017photometric,shamir2017large,shamir2017colour,shamir2020large,shamir2020pasa,shamir2021particles,shamir2021large,shamir2022new,shamir2022large,shamir2022analysis}.   

\begin{figure}[h!]
\centering
\includegraphics[scale=0.72]{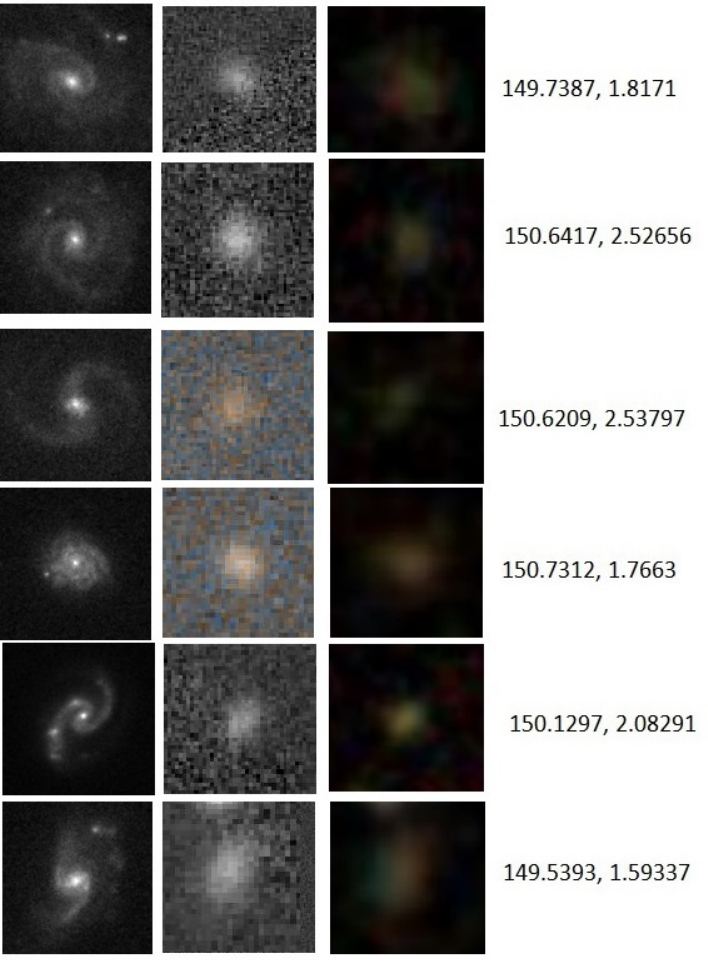}
\caption{Images of the same galaxies as imaged by HST (left column), Pan-STARRS (middle column), SDSS (right column), and the $(\alpha , \delta)$ coordinates of each galaxy. These example galaxies were taken from the COSMOS field, and are not part of the Ultra Deep Field studied in this paper.}
\label{sdss_hst}
\end{figure}

The problem of completeness is a limitation of all algorithms and all analysis methods, as the visible fine details of the shape of a galaxy depend on the quality of the imaging \citep{sym15061190,psac058}. The quality of imaging is always limited, even for powerful space-based imaging devices such as JWST and HST. For example, Figure~\ref{sdss_hst} shows images of the same galaxies imaged by SDSS and Pan-STARRS, and were also imaged by HST as part of the COSMOS survey. As the examples show, galaxies that do not seem to have an identifiable direction of rotation through SDSS and Pan-STARRS have clear spin patterns when observed through the more powerful HST. 

Therefore, a ``complete'' dataset of annotated galaxies imaged by SDSS or Pan-STARRS will be highly incomplete if the exact same galaxies were imaged by HST. Since no telescope can provide infinite imaging quality, completeness cannot be achieved. As mentioned above, the algorithm is symmetric, and therefore the galaxies that are not annotated are expected to be distributed equally, within statistical error, between galaxies that spin clockwise and galaxies that spin counterclockwise \citep{shamir2011ganalyzer,shamir2013color,shamir2016asymmetry,shamir2017photometric,shamir2017large,shamir2017colour,shamir2020large,shamir2020pasa,shamir2021particles,shamir2021large,shamir2022new,shamir2022large,shamir2022analysis}. Quantitative analysis is described in Section 4 in \cite{shamir2022analysis}. Naturally, the high quality of the JWST deep field images makes it easier to both computers and the human eye to identify the shapes of the galaxies.

The analysis method was applied to the JWST image taken at the same field as the Hubble Ultra Deep Field mentioned above, as well as to Webb's First Deep Field as will be discussed later in this paper. The method was also applied after mirroring the images. That analysis provided the same annotation, as expected due to the symmetric nature of the algorithm. Figure~\ref{jwst} shows the galaxies that were annotated by their direction of rotation. 

\begin{figure*}[h!]
\centering
\includegraphics[scale=0.17]{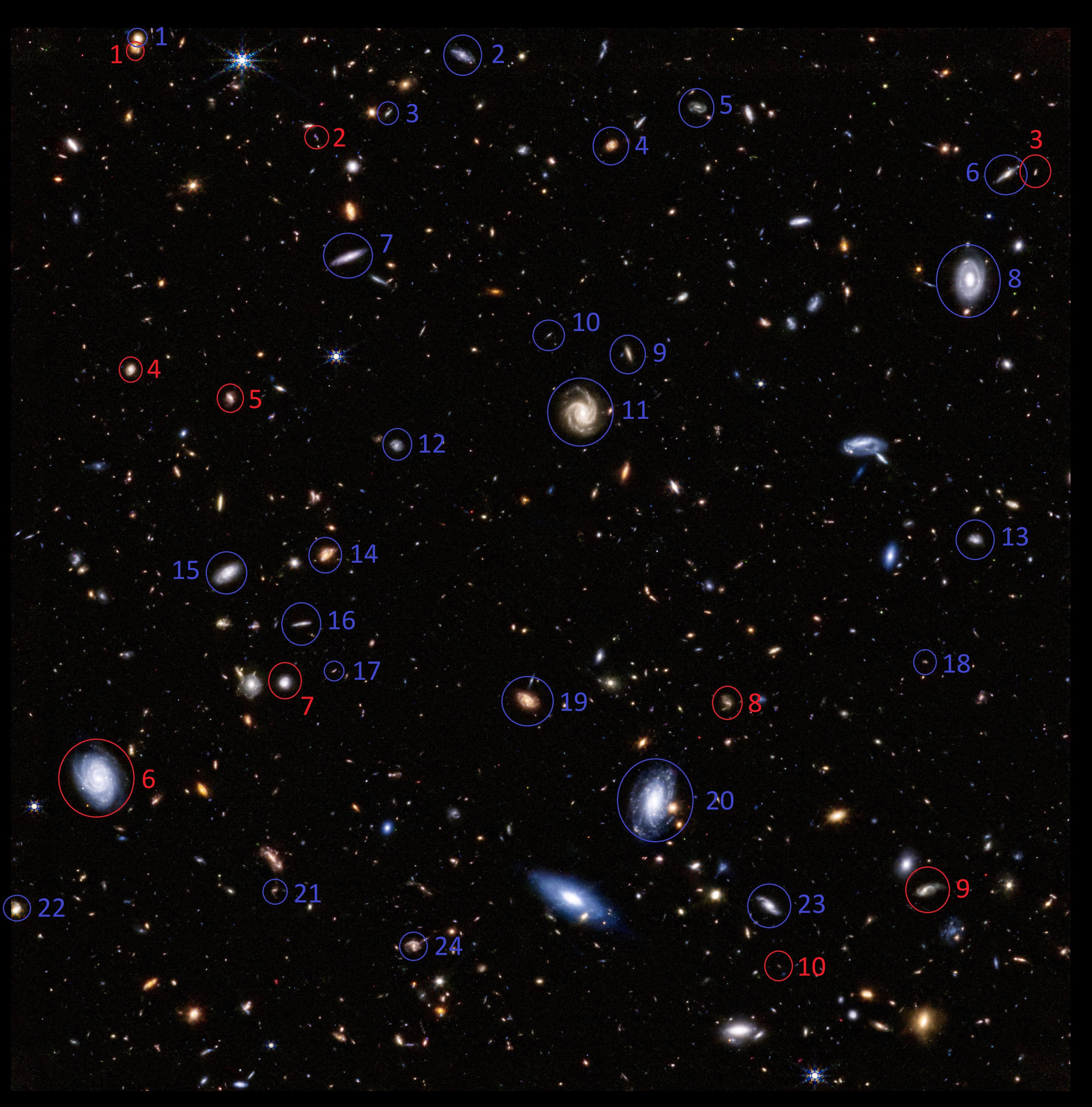}
\caption{Spiral galaxies spinning clockwise (blue) and counterclockwise (red) in JWST image taken at the same location of HST Ultra Deep Field.}
\label{jwst}
\end{figure*}


The analysis found 348 objects that met the foreground size threshold set by {\it Ganalyzer} as described above. Most of these objects did not have an identifiable spin direction, and therefore {\it Ganalyzer} did not provide annotations for most of the objects. Out of that set of objects, 34 galaxies had an identifiable direction of rotation as determined by {\it Ganalyzer}. These 34 objects were distributed such that 24 galaxies were annotated as rotating clockwise, and 10 as rotating counterclockwise. The one-tailed binomial distribution probability to have such separation or stronger by chance when assuming that the probability of a galaxy to spin in a certain direction is 0.5 is $\sim$0.012. Consequently, the two-tailed probability of the distribution is $\sim$0.024, but because previous experiments showed that a higher number of galaxies that rotate clockwise is expected in that part of the sky \citep{shamir2020large,shamir2020pasa,shamir2021particles,shamir2021large,shamir2022new,shamir2022large,shamir2022analysis}, the one-tailed statistical significance can be used. These previous experiments and their comparison to the JWST data will be discussed in Section~\ref{comparison}. The excessive number of galaxies that rotate clockwise agrees with previous reports on such distribution in the Ultra Deep Field imaged by the Hubble Space Telescope \citep{shamir2021aas}.

\begin{figure*}[h!]
\centering
\includegraphics[scale=0.8]{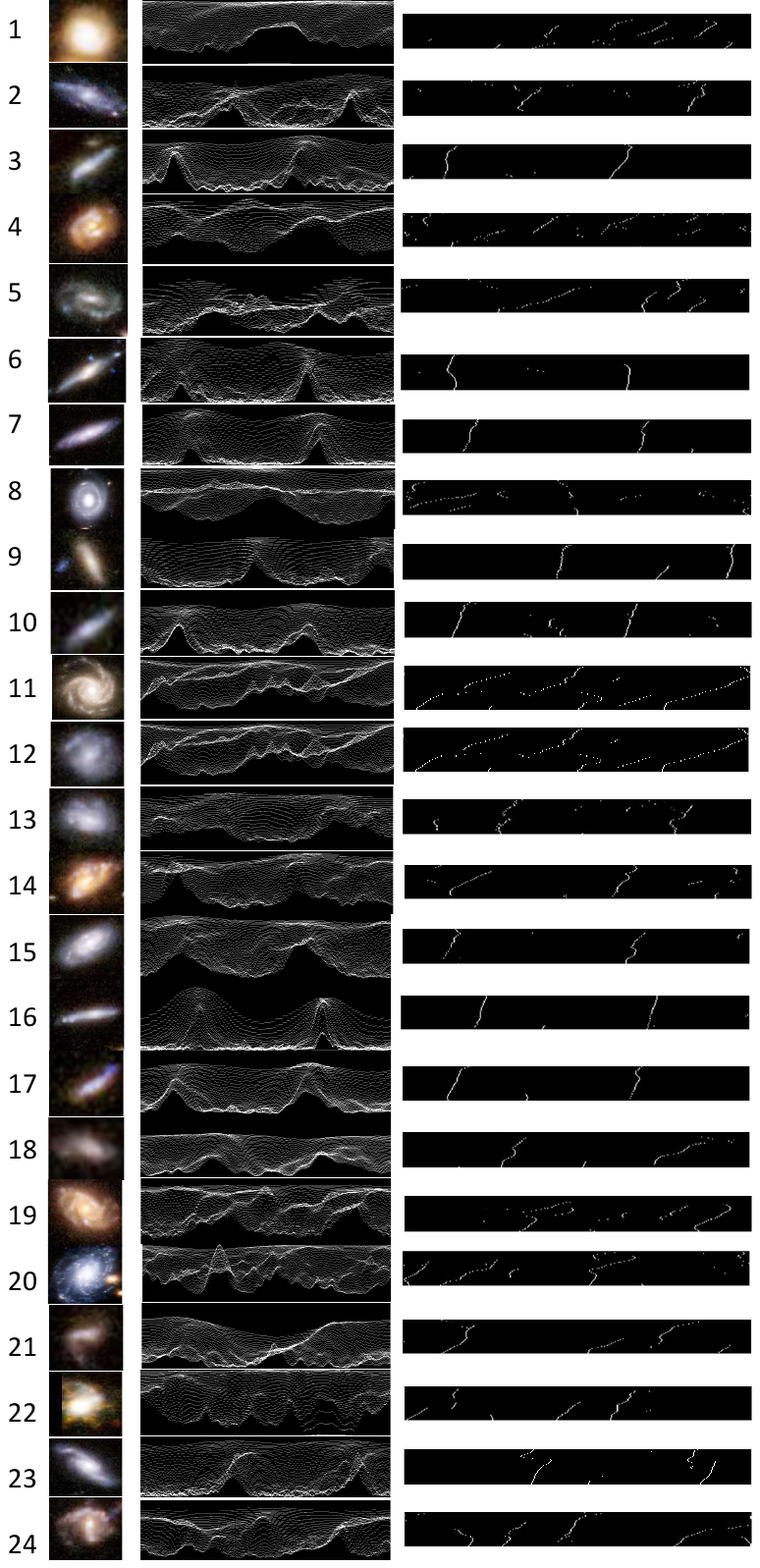}
\caption{Images of the objects that were identified as rotating clockwise (left), the radial intensity plots of each image (centre), and the peaks detected in the lines of the radial intensity plot (right).}
\label{cw}
\end{figure*}

\begin{figure*}[h!]
\centering
\includegraphics[scale=0.45]{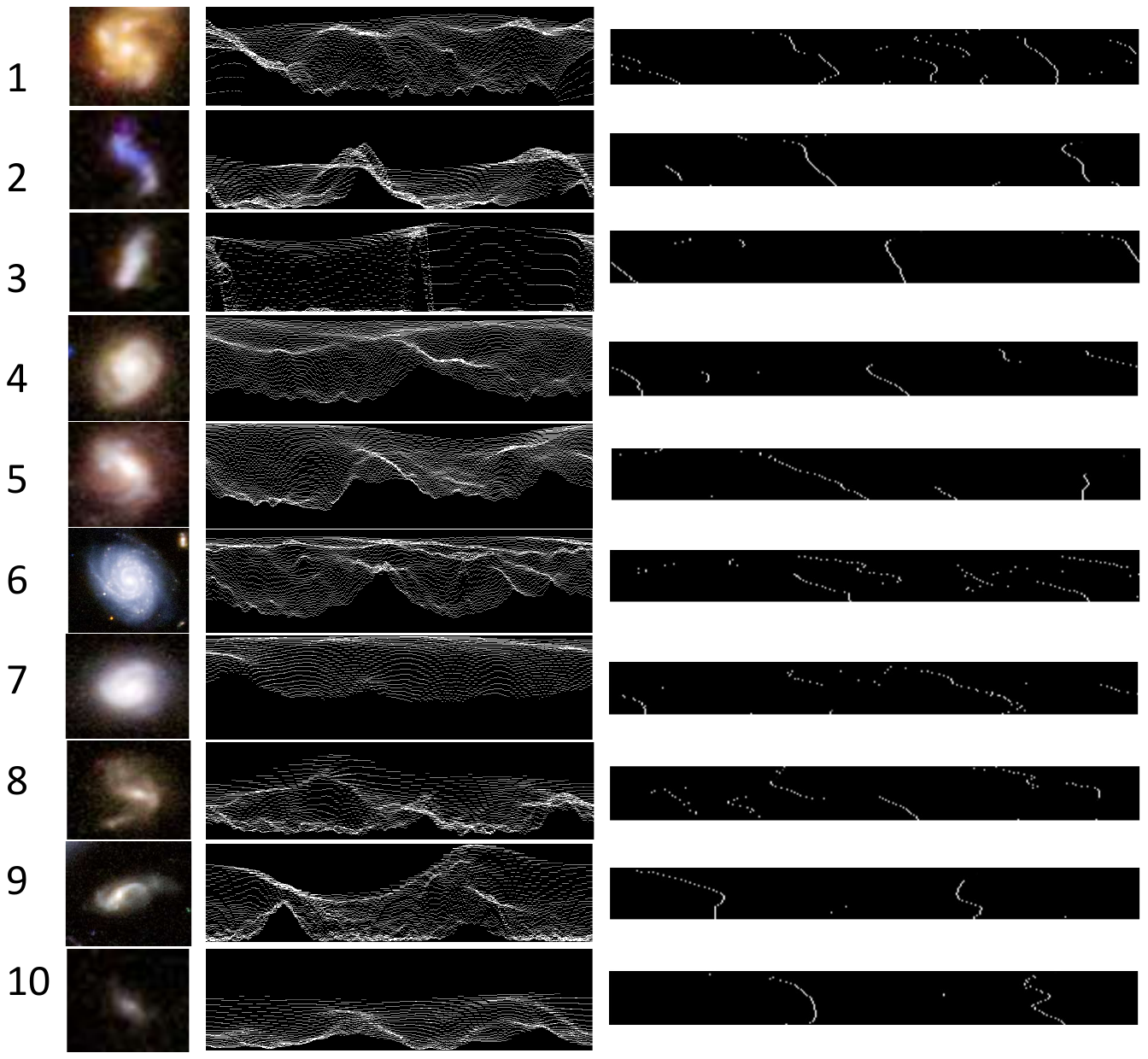}
\caption{Images of the objects that were identified as rotating counterclockwise, with the radial intensity plot of each image and the peaks detected in the lines of the radial intensity plot.}
\label{ccw}
\end{figure*}

\begin{figure*}[h!]
\centering
\includegraphics[scale=0.6]{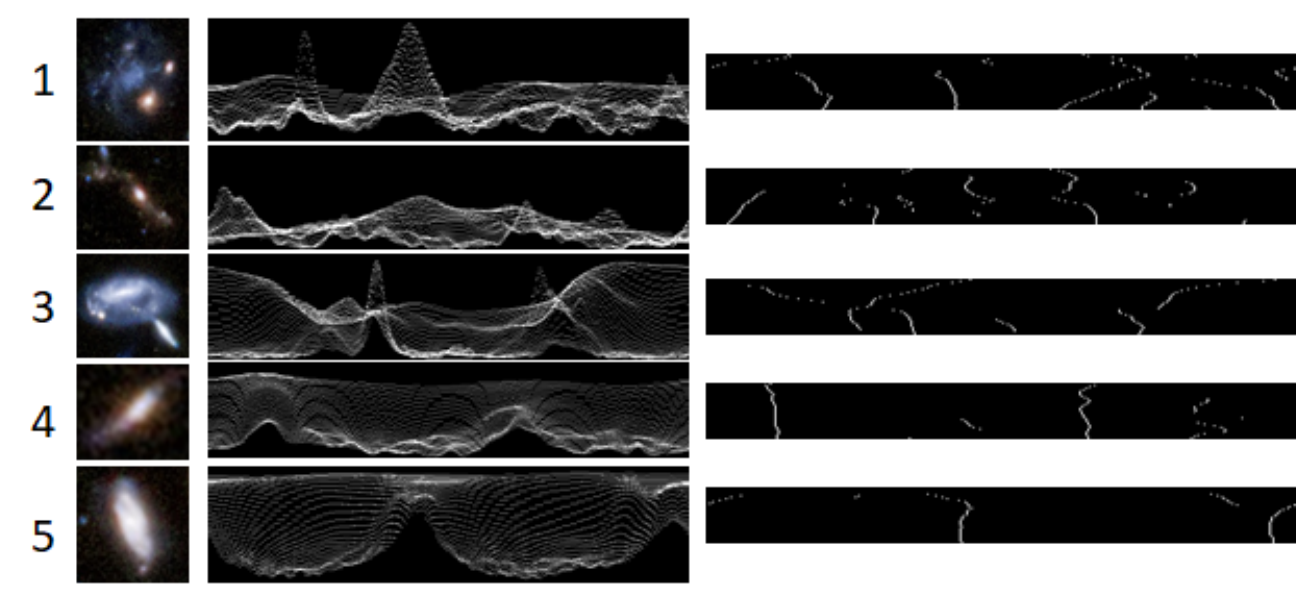}
\caption{Images of example objects that were not identified by the algorithm to have a clear direction of rotation.}
\label{undetermined}
\end{figure*}

The objects that were detected and their analyses through the radial intensity plots are shown in Figures~\ref{cw} through~\ref{undetermined}. Figure~\ref{cw} shows the extended objects identified as rotating clockwise, and Figure~\ref{ccw} shows the objects identified as rotating counterclockwise. Manual inspection of the objects suggest that their classification is aligned with manual impression, given the limitations of the human eye. Object 8 in Figure~\ref{cw} seems visually to be a ring galaxy, but it also has one trailing arm. In any case, none of the objects is missclassified in a manner that is obvious to the human eye. 

The image was also scanned manually to identify objects that might have not been identified by the algorithm but would have been identified when observed manually. Figure~\ref{undetermined} shows examples of the objects that were not identified by the algorithm to rotate towards a specific direction, but human inspection might suggest that they have a direction of rotation. It might be possible that Objects 1, 4, and 5 rotate clockwise, while Objects 3 and 4 rotate counterclockwise, but the shapes are not sufficiently clear to determine the direction of rotation. In any case, the use of the algorithm is expected to avoid bias, such that the possible bias, and the same mathematical rules that apply to galaxies that rotate clockwise are also applied to galaxies that rotate counterclockwise. Therefore, the bias of the human perception does not have any impact on the results.

\begin{figure*}[h]
\centering
\includegraphics[scale=0.75]{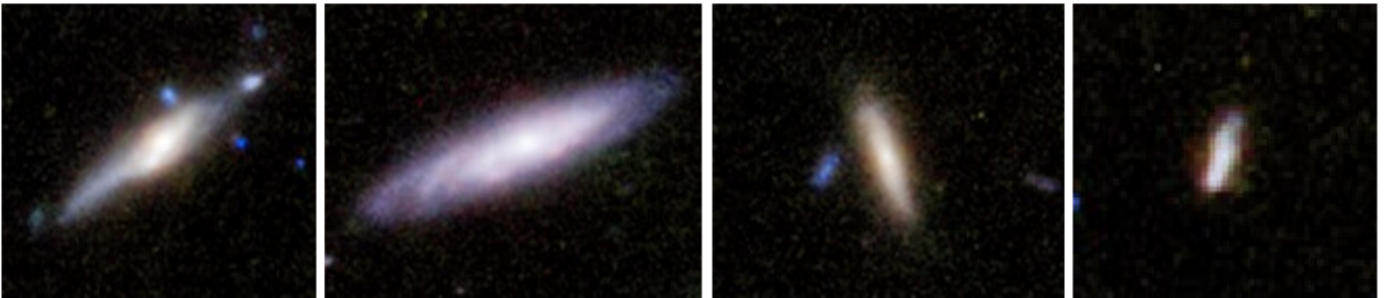}
\caption{Examples of galaxies with high inclination angle. These include galaxies 6, 7, and 9 from Figure~\ref{cw} and 3 from Figure~\ref{ccw}.}
\label{high_inclination}
\end{figure*}

Some galaxies in Figures~\ref{cw} and~\ref{ccw} have a relatively high inclination angle. Such galaxies include 6, 7, 9, 10, and 16 from Figure~\ref{cw}, and galaxies 3 and 10 from Figure~\ref{ccw}. Because of the high inclination angle these galaxies are more difficult to inspect by eye, but careful examination can reveal the galaxy arm patterns. Figure~\ref{high_inclination} shows examples of higher-resolution images of galaxies with high inclination from Figures~\ref{cw} and~\ref{ccw}.

Despite the high inclination, the arms can still be identified. Galaxy 6 has two visible arms, one on each side of the galaxy, and Galaxy 7 has several arms that their curve is noticeable in the bottom left part of the galaxy. Galaxies 9 and 3 have slight yet still visible curve of the arms that indicate on the direction of rotation. In any case, the algorithm is symmetric and applied to all galaxies in a similar manner, and all galaxies are analysed in the same way to avoid bias. The downsides of this analysis as a full proof for the presence of the asymmetry will be discussed in Section~\ref{comparison}.




\section{Comparison of the JWST observation to previous observations}
\label{comparison}

The observation discussed in Section~\ref{results} shows statistical significance of the asymmetry, but might still not be sufficient to fully prove that the number of galaxies that rotate in opposite directions as observed from Earth is indeed asymmetric. The reason is that the number of galaxies is still small. Both machines and the human eye can be sensitive to the shapes of the galaxies. Since the deep field image is a natural image, the shapes of the galaxies cannot be normalised, and the field contains a variety of different shapes of galaxies. That might theoretically lead to some galaxies being annotated inaccurately. For instance, if the machine vision or the human perception tends to annotate a certain shape as a galaxy that rotates clockwise, the presence of such shapes of galaxies in the field can bias the results. The galaxies used in Section~\ref{results} were annotated by a symmetric algorithm and were inspected by the human eye, but unknown or unexpected biases could still exist. Because any natural field is expected to contain galaxies with various shapes, an experiment that can overcome the theoretical impact of differences in the galaxy shapes needs to be based on a very large number of galaxies. When a large number of galaxies is used, the different shapes of galaxies will be distributed equally between galaxies that rotate clockwise and counterclockwise, and therefore a higher frequency of galaxy shapes in one of these classes will not be able to lead to asymmetry in the galaxy directions of rotation.
   
\begin{figure*}[h]
\centering
\includegraphics[scale=0.25]{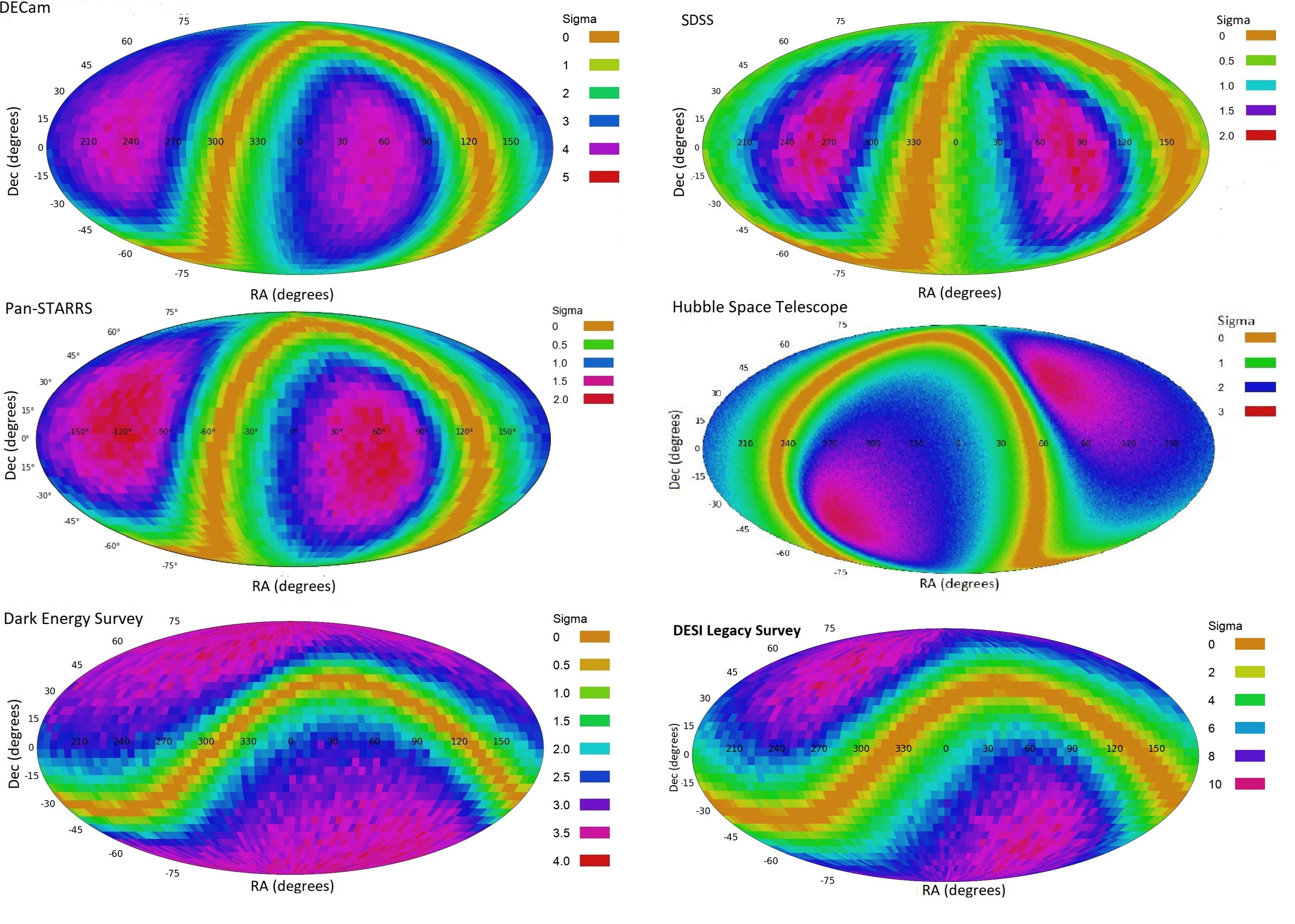}
\caption{The probability of a dipole axis formed by the asymmetry in the spin directions of spiral galaxies as determined from data collected by SDSS \citep{shamir2019large,shamir2020patterns,shamir2021particles,shamir2022large}, HST \citep{shamir2020pasa}, Pan-STARRS \citep{shamir2020patterns}, the Dark Energy Survey \citep{shamir2022asymmetry}, and DESI Legacy Survey \citep{shamir2021large,shamir2022analysis}.}
\label{decam_sdss_panstarrs_normalized}
\end{figure*}

While JWST deep fields provide a far deeper view than any Earth-based instrument, Earth-based digital sky surveys can image a far larger number of galaxies. Analysis of all existing premier digital sky surveys show statistically significant asymmetry in the distribution of galaxy spin directions across the sky, and a dipole axis formed by the distribution of galaxy spin directions that peaks around the Galactic pole. These digital sky surveys include SDSS \citep{shamir2019large,shamir2020patterns,shamir2021particles,shamir2022large}, HST \citep{shamir2020pasa}, Pan-STARRS \citep{shamir2020patterns}, the Dark Energy Survey \citep{shamir2022asymmetry}, and DESI Legacy Survey \citep{shamir2021large,shamir2022analysis}, all of them show very similar results \citep{shamir2022large}. Figure~\ref{decam_sdss_panstarrs_normalized} shows the probability that a dipole axis is formed from the galaxy spin directions by mere chance at different parts of the sky. Several analyses are shown with data from several different sky surveys, as discussed thoroughly in \citep{shamir2019large,shamir2020patterns,shamir2020pasa,shamir2021large,shamir2021particles,shamir2022large,shamir2022asymmetry,shamir2022analysis}.  

As the figure shows, all digital sky surveys provide comparable results regarding the most likely position of the dipole axis formed by the galaxy spin directions. The location of the Ultra Deep Field is in the part of the sky where the number of galaxies that rotate clockwise is higher. These previous reports also provide information about the higher number of galaxies rotating clockwise in the Southern hemisphere, and specifically around the Ultra Deep Field.


Table~\ref{decam} shows the distribution of the galaxies in DESI Legacy Survey in the field $(38.15^o<\alpha<68.15^o,-42.78^o<\delta<-12.78^o)$, which is the $30^o\times30^o$ sky around the location of the Ultra Deep Field. The data are taken from the dataset of $\sim1.3\cdot10^6$ DESI Legacy Survey galaxies separated into clockwise and counterclockwise \citep{shamir2022analysis}. As the table shows, the data from DESI Legacy Survey shows a higher number of galaxies that rotate clockwise around the Ultra Deep Field. The footprint of the DESI Legacy Survey is sufficiently large to cover both ends of the galactic pole, and therefore the asymmetry can be compared to the asymmetry in the corresponding field in the opposite hemisphere. The asymmetry in the corresponding field in the opposite hemisphere has a lower number of galaxies, and the asymmetry is not statistically significant, but it shows a higher number of galaxies that rotate counterclockwise.

\begin{table*}
\centering
\small
\begin{tabular}{lcccc}
\hline
Field & \# cw    & \# ccw    & $\frac{cw-ccw}{cw+ccw}$  & P \\
        & galaxies & galaxies  &                          &   \\
\hline
$(38.15^o<\alpha<68.15^o,-42.78^o<\delta<-12.78^o)$    &   29,447    & 28,870  &   0.01      &  0.008  \\   
$(218.15^o<\alpha<248.15^o,12.78^o<\delta<42.78^o)$  &   15,242    & 15,356  &   -0.0037    &  0.25   \\
\hline
\end{tabular}
\caption{Distribution of clockwise and counterclockwise galaxies in the field of $(43.15^o<\alpha<63.15^o,-37.78^o<\delta<17.78^o)$ in the DESI Legacy Survey, and in the corresponding field in the opposite hemisphere. The P value is the binomial distribution probability to have such difference or stronger by chance when assuming 0.5 probability for a galaxy to spin clockwise or counterclockwise. }
\label{decam}
\end{table*}

While analysis of the data from DESI and JWST provides agreement on the existence and direction of the asymmetry, the asymmetry observed in DESI Legacy Survey is far lower than the asymmetry observed in JWST. As shown in \cite{shamir2020patterns,shamir2022large}, the asymmetry gets stronger as the redshift gets higher. The JWST imaging is sensitive to galaxies at very high redshifts, including those at higher redshift than in previous work. If the previously established trend of increasing asymmetry at greater redshift \citep{shamir2020patterns,shamir2022large} continues, that would be consistent with the degree of asymmetry seen in the current analysis.

\begin{figure*}[h]
\centering
\includegraphics[scale=0.4]{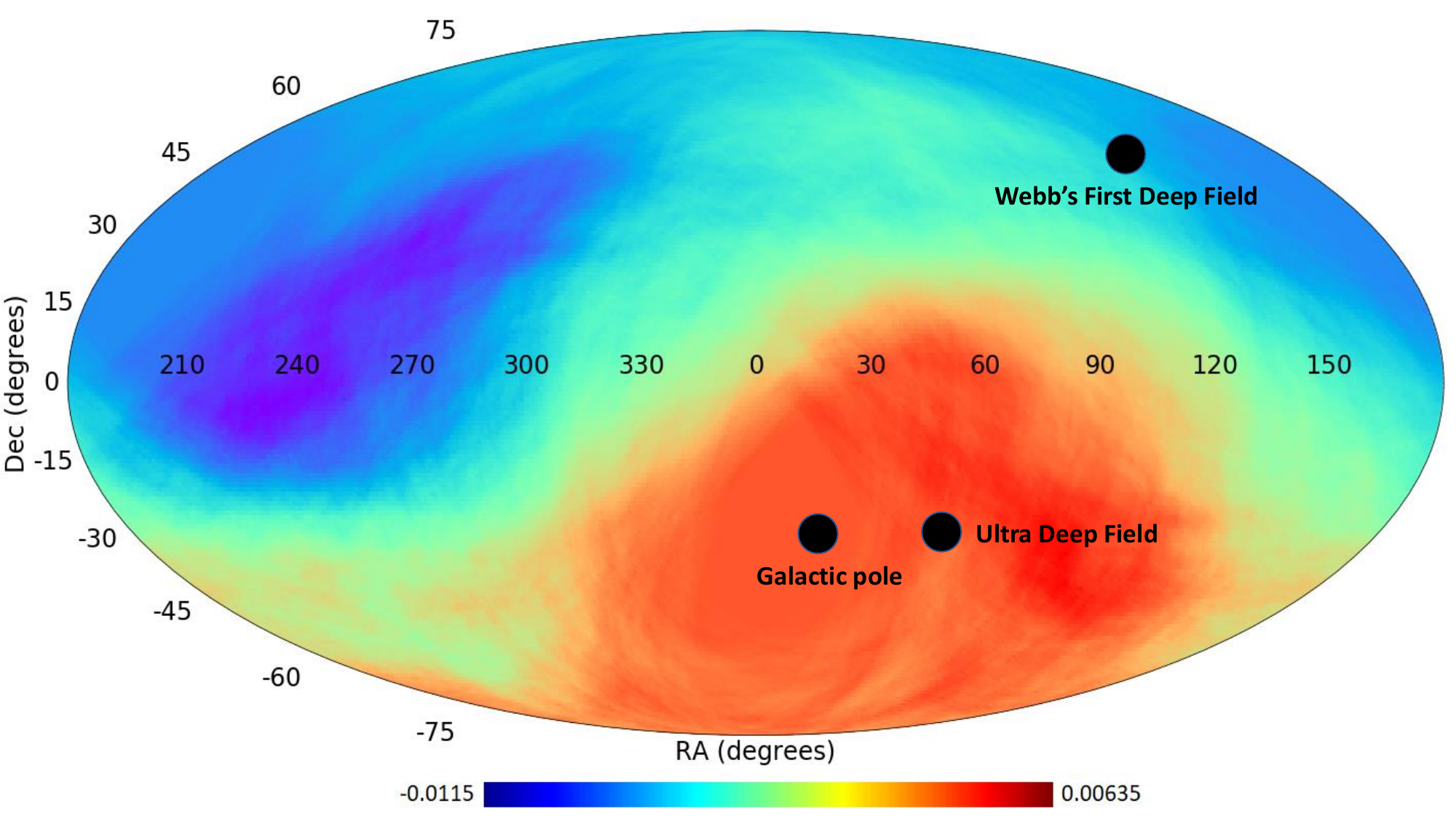}
\caption{The asymmetry in the distribution of the spin directions of $1.3\cdot10^6$ DESI Legacy Survey galaxies in different parts of the sky \citep{shamir2022analysis}. The location of the Ultra Deep Field shows that the excessive number of galaxies that rotate clockwise in that field was expected from previous analysis using Earth-based sky surveys.}
\label{jwst_location}
\end{figure*}

Figure~\ref{decam_sdss_panstarrs_normalized} shows the results of statistical analysis of the fitting the galaxy spin directions to a dipole axis \citep{shamir2019large,shamir2020patterns,shamir2020pasa,shamir2021large,shamir2021particles,shamir2022large,shamir2022asymmetry,shamir2022analysis}. Given a large number of galaxies, the asymmetry in galaxy spin directions in different parts of the sky can be inferred based on direct measurement, rather than fitting to a statistical model. Figure~\ref{jwst_location} shows the asymmetry in different parts of the sky using $1.3\cdot10^6$ DESI Legacy Survey galaxies annotated by their direction of rotation. The analysis is described in detail in \cite{shamir2022analysis}, and shows the ratio between galaxies that rotate in opposite directions in different fields in the sky \citep{shamir2022analysis}. Red parts of the sky are areas with an excessive number of galaxies that rotate clockwise, while blue parts show parts of the sky where the number of galaxies that rotate counterclockwise is higher. The figure also shows the location of the Ultra Deep Field. As can be seen in the figure, the previous analysis, like the observations shown in Figure~\ref{decam_sdss_panstarrs_normalized}, shows a higher number of galaxies that rotate clockwise in that part of the sky, which is aligned with the distribution of the galaxies as observed in the JWST image of that field. As shown previously \citep{shamir2021aas}, the HST Ultra Deep Field also shows a higher number of galaxies rotating clockwise.

Figure~\ref{jwst_location} also shows that the magnitude of the asymmetry observed in DESI Legacy Survey at around the field of the Ultra Deep Field is much smaller than the magnitude of the asymmetry discussed in Section~\ref{results}. That is aligned with the previous observation that the magnitude of the asymmetry becomes stronger as the redshift gets higher \citep{shamir2019large,shamir2020patterns,shamir2022large}. For instance, Tables 3, 5, 6 and 7 in  \cite{shamir2020patterns} show that trend, as does Figure 7 in \cite{shamir2022large}. These all show that the magnitude of the asymmetry grows consistently as the redshift of the galaxies increases. If these observations are an accurate representation of structure in the Universe, they mean that in the earlier Universe spiral galaxies were more likely to rotate in the same direction, and the direction of rotation of spiral galaxies became more random over time. That observation agrees with the possible stronger asymmetry observed in a deep field image acquired by the far more sensitive JWST.

One of the most studied JWST fields acquired to date is the ``Webb's First Deep Field'' (SMACS J0723.3-7327). A similar analysis to the analysis described in Section~\ref{results} led to 19 galaxies that rotate clockwise, and 21 galaxies that rotate counterclockwise. That difference is not significant, which agrees with the isotropy assumption of the Universe. As Figure~\ref{jwst_location} shows, Webb's First Deep Field is at a part of the sky where weak asymmetry in the distribution of galaxy spin directions is expected.

The field taken by JWST shown in Figure~\ref{jwst} does not yet have redshifts for the galaxies, and therefore the exact redshifts of these galaxies are not known. But examining the redshift of some of the galaxies in the HST Ultra Deep Field (UDF), the redshift of these galaxies is far higher than in any Earth-based digital sky survey. For instance, an analysis of 16 galaxies whose morphology is clear in the HST UDF showed that the average redshift was 2.13, and the lowest redshift was 0.66 \citep{dunlop2017deep}. Because the JWST deep field is expected to be at least as deep as HST deep field, it can be reasonably assumed that the reshifts of the galaxies in Figure~\ref{jwst} are, on average, at least as high, and therefore far higher than galaxies imaged by Earth-based digital sky surveys.
 
\begin{figure}[h]
\centering
\includegraphics[scale=0.66]{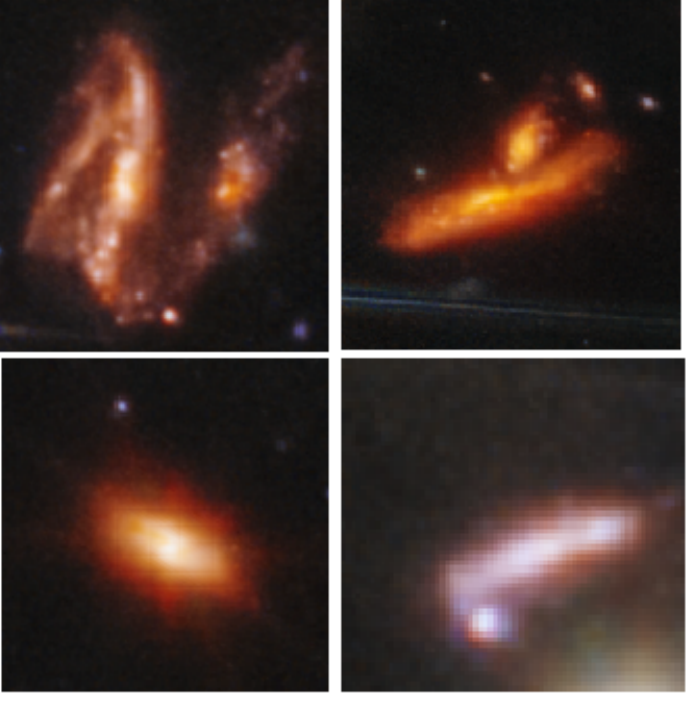}
\caption{Examples of galaxies from Webb's First Deep Field. All of these galaxies are part of an overdensity of galaxies in that field, and their redshift is 1.97. Despite the high redshift, the details of the shapes of the galaxies are still clear.}
\label{first_deep_field_examples}
\end{figure}

Also, the experience from Webb's First Deep Field showed that JWST is able to image galaxies at high redshifts with excellent image quality allowing for analysis of their shapes. For instance, Figure~\ref{first_deep_field_examples} shows four galaxies that are part of an overdensity of galaxies in Webb's First Deep Field that have redshift of 1.97 \citep{noirot2023first}. As the figure shows, despite the relatively high redshift, the shapes of these galaxies are still clear and detailed.

\section{Summary of experiments that showed random distribution of spin directions of spiral galaxies}
\label{other_reports}

Sections~\ref{introduction} and~\ref{comparison} mention several previous reports that argue that the number of galaxies that rotate in opposite directions is not necessarily symmetric, and therefore in agreement with the observation described in Section~\ref{results}. On the other hand, several other previous studies suggested no asymmetry \citep{iye1991catalog,land2008galaxy,hayes2017nature,tadaki2020spin,iye2020spin}, and are therefore in conflict with the results shown in Section~\ref{results} and in the studies mentioned in Section~\ref{comparison}. 
Detailed re-analysis of those studies using the same data finds that all are in full agreement with the contention of an asymmetric distribution, as also found here. Full details of the reproduction and analysis, including code and data to reproduce the experiments are available in \cite{sym15091704}, and discussions and reproduction experiments are also available in previous reports \citep{psac058,mcadam2023reanalysis,shamir2022analysis}.

In summary, several experiments were done, with different methodologies to annotate the galaxies. An early experiment \citep{iye1991catalog} used manual annotation of the galaxies. The initial dataset included 8,297 spiral galaxies taken from the ESO/Uppsala Survey of the ESO(B) Atlas. The survey acquired images from the Southern hemisphere ($\delta<-20$) using telescopes in La Silla, Chile, and Siding Spring, Australia. The manual analysis of the galaxy images found 3,257 galaxies that rotate clockwise and 3,268 galaxies that rotate counterclockwise. As shown quantitatively by statistical analysis  \citep{sym15091704,psac058,mcadam2023reanalysis,shamir2022analysis}, the number of galaxies analysed in that experiment is far too small to show a statistically significant asymmetry with galaxies of relatively low redshift, and therefore statistically significant asymmetry is not expected in that relatively small manually annotated dataset.

A highly publicized experiment with a larger number of galaxies was performed by using SDSS as a source for the data, and crowdsourcing for annotating the galaxies through a web-based user interface \citep{land2008galaxy}. The combination of a digital sky survey with crowdsourcing could provide a solution to the annotation of a large number of galaxies, which consequently could provide an answer to the distribution of spiral galaxies as observed from Earth. The major downside of the experiment was that the annotations were heavily biased by the human perception of the anonymous non-scientist annotators, and therefore the vast majority of the galaxies that were annotated could not be used, as the annotations of the different annotators conflicted with each other. More importantly, only the original images of the galaxies were annotated, and the images were not mirrored to offset for a human or user-interface bias. That led to a very strong asymmetry in the data of $\sim$15\%, even when using just the most clean ``superclean'' data, in which only galaxies that the annotators agreement was 95\% or higher were used. The very strong asymmetry driven by the human perceptual bias did not allow identification of a possible, likely smaller, asymmetry if such indeed existed in the real sky. 

When the problem was noticed, another experiment was performed, in which the images were annotated also after mirroring the images. Annotating both the original and mirrored images ensured that the perceptional bias in the annotations of each image was offset by the annotations of the mirrored image. But because the problem was identified after many galaxies were already annotated, the experiment with the mirrored galaxies included a relatively small number of 91,303 galaxies, that eventually led to a dataset of $\sim1.1\cdot10^4$ annotated galaxies shown in Table 2 in \cite{land2008galaxy}. That paper indicates that the data does not show statistically significant asymmetry between galaxies annotated by their direction of rotation, yet without providing a statistical analysis or P values. As shown in previous reports \citep{sym15091704,shamir2022analysis,mcadam2023reanalysis}, the asymmetry of the numbers reported in Table 2 in \cite{land2008galaxy} is in very good agreement with the asymmetry identified by the analysis of SDSS galaxies in the same footprint \citep{shamir2020patterns}.  

According to Table 2 in \cite{land2008galaxy}, 5.525\% of the galaxies were annotated as rotating clockwise, and 5.646\% of the mirrored galaxies were annotated as rotating clockwise. That shows that 2.2\% more galaxies rotate counterclockwise. Similarly, 6.032\% were annotated as rotating counterclockwise in the original images, compared to just 5.942\% of the galaxies that were annotated as rotating counterclockwise in the mirrored images, showing that 1.5\% more galaxies rotate counterclockwise. As explained in \cite{sym15091704,psac058,shamir2022analysis}, the 1.5\%-2.2\% asymmetry is in excellent agreement with the asymmetry reported in \cite{shamir2020patterns}, which was analysed using the same footprint of SDSS galaxies with redshift. When combining the P values of the two experiments, the asymmetry becomes statistically significant \citep{sym15091704,psac058,mcadam2023reanalysis}. Table~\ref{galaxy_zoo_table} in this paper shows the number of galaxies that rotate in opposite directions used in the two experiments described by \cite{land2008galaxy} when the galaxy images were mirrored to offset for the human bias.

Like \cite{land2008galaxy}, the experiment reported in \cite{shamir2020patterns} used SDSS galaxies with redshift, and therefore the footprints were similar in both experiments. The main difference between the experiments was the number of galaxies used. The experiment described in \cite{shamir2020patterns} used over $6\cdot10^4$ galaxies, enabling a stronger statistical signal. But as also shown in \cite{sym15091704,psac058}, the statistical signal of the data shown in Table 2 in \cite{land2008galaxy} is not necessarily statistically insignificant. 


\begin{table}
\centering
\scriptsize
\begin{tabular}{|l|c|c|c|c|}
\hline
                   & Original          & Mirrored   & $\frac{\#ccw}{\#cw}$        & P                      \\
                   &                      &                &                                    & (one-tailed)       \\   
\hline
Clockwise             & 5,044      & 5,155       &   1.022                          &    0.13   \\     
Counterclockwise  &  5,507     & 5,425      &   1.015                          &    0.21    \\     
\hline
\end{tabular}
\caption{The number of galaxies rotating in opposite directions in the original and mirrored images used in \cite{land2008galaxy}.}
\label{galaxy_zoo_table}
\end{table}

Another experiment \citep{hayes2017nature} that used SDSS galaxies with redshift used computer analysis to annotate the direction of rotation of the galaxies that were annotated manually by \cite{land2008galaxy}. That experiment is analysed and reproduced in \cite{mcadam2023reanalysis}, and discussed also in \cite{sym15091704}. As shown by Table 2 in \cite{hayes2017nature}, the analysis showed a higher number of galaxies rotating counterclockwise, with statistical significance of between 2$\sigma$ to 3$\sigma$. The results of these experiments are in agreement with the results shown in \cite{shamir2020patterns}, that also annotated galaxies in the same footprint, and also showed a higher number of galaxies rotating counterclockwise. These experiments were based on crowdsourcing selection of the spiral galaxies, before annotating the spiral galaxies automatically using machine vision. Therefore, the results could have been subjected to a bias that was not known previously, which is a possible human bias in the separation between spiral and elliptical galaxies. 

To avoid such human bias, another experiment was performed such that the spiral galaxies were separated from the elliptical galaxies by using a machine learning method. That analysis provided random distribution. But a careful analysis of the process showed that the machine learning method was used after removing manually all features that were correlated with the asymmetry in the distribution of galaxy spin directions. As described by \cite{hayes2017nature}, ``We choose our attributes to include some photometric attributes that were disjoint with those that Shamir (2016) found to be correlated with chirality, in addition to several SpArcFiRe outputs with all chirality information removed''. The paper does not provide a motivation for manually removing just the features that are associated with the asymmetry.

\begin{figure}
\includegraphics[scale=0.24]{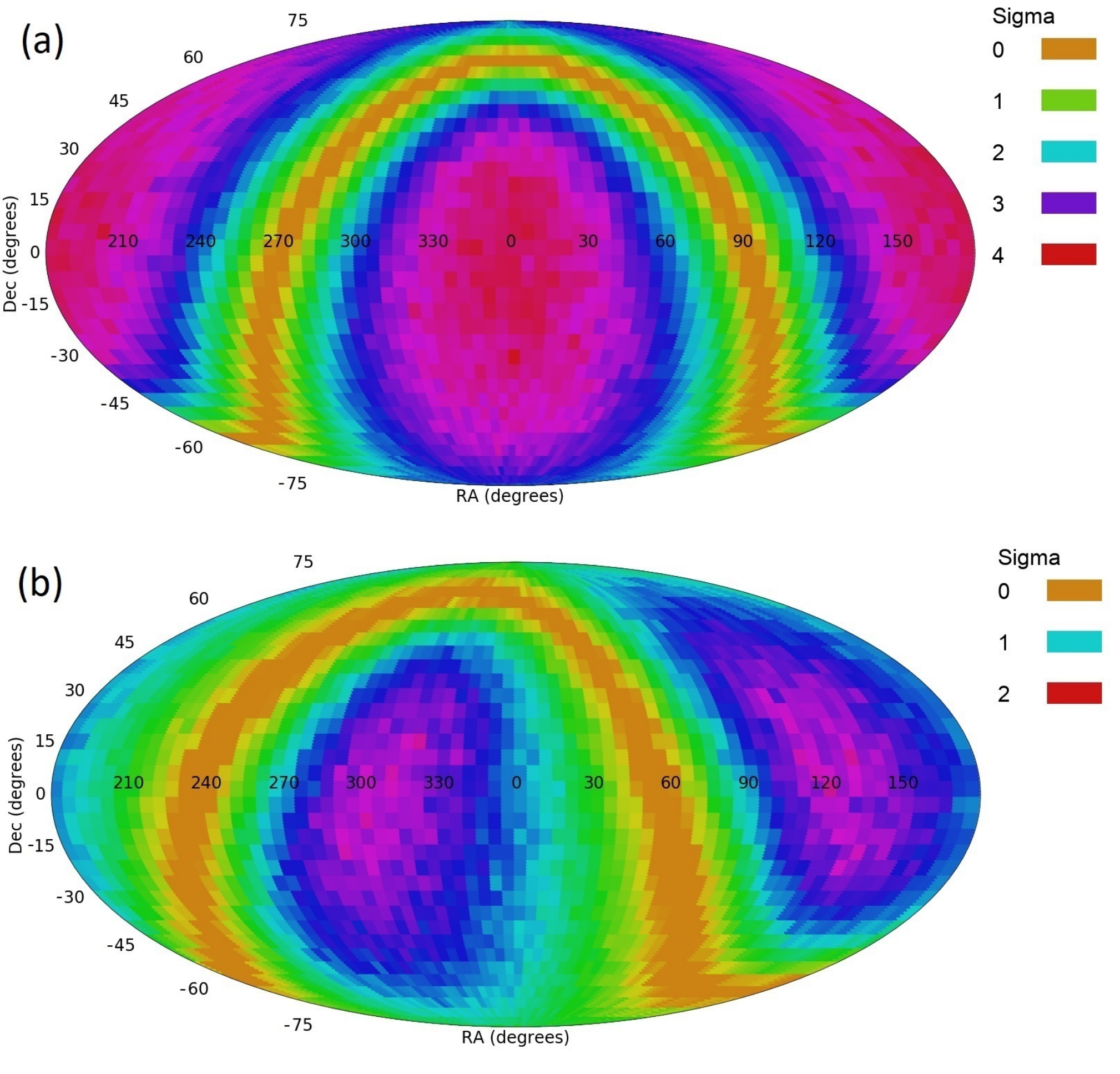}
\caption{Results of analysis of SDSS galaxies used in \cite{hayes2017nature}. Panel (a) shows the result of the analysis after separating the spiral galaxies from the elliptical galaxies before annotating them by their direction of rotation. Panel (b) shows the results of the analysis without applying a first step of separating spiral galaxies from the rest of the galaxies.  The full description of these experiments, with code and data to reproduce them, are provided in \cite{mcadam2023reanalysis}. }
\label{sparcfire}
\end{figure}

Naturally, when manually removing just the features that correlate with the asymmetry, the asymmetry is weakened. Reproduction of the same analysis of \cite{hayes2017nature} with the same code and the same data, but without manually removing features is shown in \cite{mcadam2023reanalysis}. Figure~\ref{sparcfire} shows the asymmetry. The experiments are done when selecting spiral galaxies before annotating them as done by \cite{hayes2017nature}, but without removing manually any features. In other experiments no selection of spiral galaxies was applied. 
In all cases the results were statistically significant. When applying a first step of automatic selection of spiral galaxies before annotating them the statistical significance was 3.6$\sigma$. When not applying any selection of the spiral galaxies the statistical significance was 2.05$\sigma$. The full description of these experiments, with code and data to reproduce them, are provided in \cite{mcadam2023reanalysis}.

Another experiment used deep convolutional neural networks to annotate a large number of Hyper Suprime-Cam (HCS) galaxies by their direction of rotation \citep{tadaki2020spin}. That experiment provided 38,718 galaxies that rotate clockwise, and 37,917 galaxies that rotate counterclockwise \citep{tadaki2020spin}. Based on the binomial distribution, the one-tailed probability of the asymmetry to occur by chance is P=0.0019. The excessive number of galaxies rotating clockwise is also in agreement with previous analyses regarding the part of the sky around the HCS footprint \citep{shamir2021large,shamir2022analysis}, as also explained in \cite{mcadam2023reanalysis}. The \cite{tadaki2020spin} analysis still suggests that the asymmetry is not statistically significant. The insignificance of the results, however, is not due to the statistical inference of the outcomes, but due to the bias of convolutional neural networks. Convolutional neural networks can have complex biases that are very difficult to profile and control \citep{DHAR2022100545,DHAR202192,ball2023}, and therefore deep neural networks might not be a sound method for this task. But as discussed in \cite{sym15091704}, while neural networks might not be a method that can be fully trusted, the results of the neural network are certainly not in conflict with the analysis shown in Section~\ref{comparison}, and in fact in agreement with these observations.

Deep neural networks were also applied by \cite{jia2023galaxy} to galaxy images from SDSS and DESI. Because the deep neural network classification has a certain degree of error, the experiment used different thresholds of labeling certainty to balance between the number of galaxies and the accuracy of the labeling.  When using the most accurate labeling threshold of 0.9, the analysis provided 9,218 SDSS galaxies spinning clockwise and 9,442 SDSS galaxies spinning counterclockwise, as shown in Table 1 in \cite{jia2023galaxy}. That provides an asymmetry of $\sim$2.4\%, which is comparable to the asymmetry shown in \cite{shamir2020patterns} or in Table~\ref{galaxy_zoo_table} for the same sky survey and the same footprint of SDSS galaxies with spectra. The one-tailed probability to have such asymmetry by chance is $\sim$0.05, which is weaker than the probability shown in \cite{shamir2020patterns}, possibly due to the lower number of galaxies. When using the DESI galaxies, the most accurate analysis provided 11,649 and 11,919 galaxies rotating clockwise and counterclockwise, respectively. The one-tailed probability of the asymmetry to occur by chance is $\sim$0.04. The lower statistical significance compared to the analysis of DESI galaxies \citep{shamir2022analysis} can be explained by the far higher number of galaxies used in \cite{shamir2022analysis}, which exceeds $10^6$ galaxies. Also, DESI covers a very large footprint that includes both hemispheres. Because the asymmetry in opposite hemispheres is inverse \citep{shamir2022analysis}, combining galaxies from the two hemispheres might weaken the signal, as the asymmetry in one hemisphere might offset the inverse asymmetry in the opposite hemisphere when the galaxies from both hemispheres are combined.

Another experiment that proposed that the number of galaxies rotating in opposite directions is equal is \citep{iye2020spin}, claiming that the asymmetry observed in previous experiment is due to ``duplicate objects'' in the dataset. This experiment is discussed in detail with replication of the analysis in \cite{psac058,sym15091704,shamir2022analysis}. In summary, the dataset used by \cite{iye2020spin} was taken from \citep{shamir2017photometric}, which was a dataset used for photometric analysis, and no claims for any kind of dipole axis formed by the galaxies in that dataset was made in the \cite{shamir2017photometric} analysis. When using the data used in \cite{shamir2017photometric} to analyse the number of galaxies rotating in opposite directions, photometric objects that are part of the same galaxy indeed become ``duplicate objects'', but as stated above, no claim for any dipole axis of any kind was made in \cite{shamir2017photometric}, and no such claim about that dataset was made in any other paper.

As explained in detail in \cite{psac058}, the analysis shown by \cite{iye2020spin} was a three-dimensional analysis according to which the location of each galaxy was determined by its right ascension, declination, and redshift. But because the galaxies used in \cite{shamir2017photometric} do not have spectra, the analysis was based on the photometric redshift. The photometric redshift is highly inaccurate, and therefore the use of inaccurate data weakens the statistical signal \citep{psac058}. 

\begin{figure*}
\centering
\includegraphics[scale=0.45]{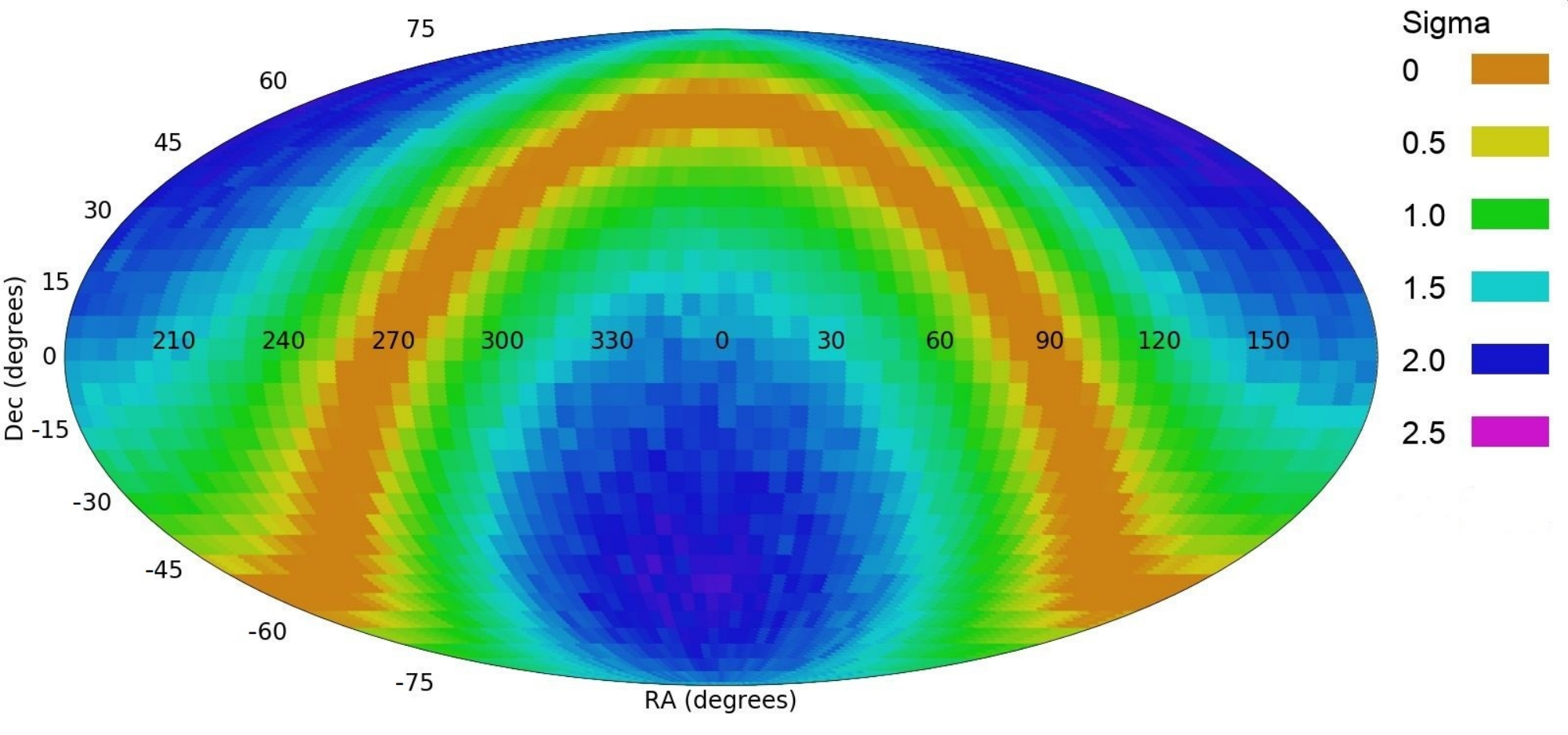}
\caption{Results of the reproduction of \citep{iye2020spin}. Full description as well as code and data to reproduce the analysis is available in \cite{sym15091704}. }
\label{dipole1}
\end{figure*}

More importantly, as discussed in detail in \cite{sym15091704}, reproducing the experiment with the same data and same analysis described in \cite{iye2020spin} provides completely different results than the results shown by \cite{iye2020spin}. Figure~\ref{dipole1} shows the outcome of the experiment, showing a statistically significant dipole axis of 2.14$\sigma$ formed by the galaxy spin directions. The full code and data to reproduce the experiment is publicly available\footnote{\label{note1}\url{https://people.cs.ksu.edu/~lshamir/data/iye_et_al}}. The explanation of the National Astronomical Observatory of Japan (NAOJ) for the differences between the results shown by \cite{iye2020spin} and the reproduction of the analysis is discussed in \cite{sym15091704}, and can also be found with the data and code\footnotemark[3].  

\section{Discussion and conclusions}
\label{conclusion}

Asymmetry between the number of galaxies rotating clockwise and the number of galaxies rotating counterclockwise in a given field and across different parts of the sky have been observed previously with Earth-based telescopes such as SDSS \citep{longo2011detection,shamir2012handedness,shamir2013color,shamir2016asymmetry,shamir2017photometric,shamir2017large,shamir2017colour,shamir2020large,shamir2020pasa,shamir2021particles,shamir2022new,shamir2022asymmetry,shamir2022analysis}, Pan-STARRS \citep{shamir2019large,shamir2020patterns}, and DECam \citep{shamir2021large,shamir2022new}. It has also been shown though observations made by space-based instruments such as HST \citep{shamir2020pasa,shamir2020asymmetry,shamir2021aas}. Analysis with large number of galaxies shows that the galaxy spin directions form a cosmological-scale dipole axis, and such axis is consistent across all premier digital sky surveys \citep{shamir2022asymmetry,shamir2022analysis}.

Despite these previous observations, the asymmetry between clockwise and counterclockwise spiral galaxies might be considered unexpected, and a possible shift from the standard cosmological model
\citep{turner1996cosmology,pecker1997some,perivolaropoulos2014large,bull2016beyond,velten2020hubble,netchitailo2020world}. While the Cosmological Principle is a common working assumption, the Cosmological Principle and the standard cosmological model have been questioned for several decades \citep{kroupa2012dark}. For instance, Jean-Claude Pecker proposed substantial observational evidence against the standard model \citep{pecker1997some}. Such observations can be related to a large number of probes that are in disagreement with the Cosmological Principle. A detailed summary of these probes is available in \cite{Aluri_2023}.

While not necessarily fully aligned with $\Lambda$CDM, the contention of a cosmological-scale dipole axis agrees with several other existing theories. These cosmological models can be related to the geometry of the Universe such as the ellipsoidal Universe 
\citep{campanelli2006ellipsoidal,campanelli2007cosmic,campanelli2011cosmic,gruppuso2007complete,cea2014ellipsoidal}, dipole big bang \citep{allahyari2023big,krishnan2023dipole}, or isotropic inflation 
\citep{arciniega2020geometric,edelstein2020aspects,arciniega2020towards,jaime2021viability,feng2003double,piao2004suppressing,bohmer2008cmb,luongo2022larger,dainotti2022evolution}.

Another cosmological model that assumes the existence of a cosmological-scale axis is the model of a rotating Universe \citep{godel1949example}. While early models of rotating universe conflict with the concept of inflation \citep{godel1949example}, more recent theories have expanded the model to also include  inflation 
\citep{ozsvath1962finite,ozsvath2001approaches,sivaram2012primordial,chechin2016rotation,seshavatharam2020integrated,camp2021}.

The rotating Universe model is also closely related to the theory of black hole cosmology
\citep{pathria1972universe,stuckey1994observable,easson2001universe,chakrabarty2020toy,tatum2018flat}.
Since black holes spin
\citep{mcclintock2006spin,mudambi2020estimation,reynolds2021observational}, a Universe hosted in a black hole is expected to spin. It has therefore been proposed that a Universe hosted in a black hole should inherit the preferred spin direction of the black hole
  \citep{poplawski2010cosmology,seshavatharam2010physics,christillin2014machian,seshavatharam2020light,seshavatharam2020integrated}.
Black hole cosmology is also related to the holographic universe \citep{susskind1995world,bak2000holographic,bousso2002holographic,myung2005holographic,hu2006interacting,rinaldi2022matrix}, which can represent the large-scale structure of the Universe in a hierarchical manner \citep{sivaram2013holography,shor2021representation}. 

On the other hand, the observation shown here through JWST as well as other telescopes as described above can also be related to the internal structure of galaxies and the physics of galaxy rotation, and not necessarily to the large-scale structure of the Universe. As also mentioned in Section~\ref{comparison}, the most likely dipole axis formed by the galaxy spin directions is close to the Galactic pole. While that can be considered a coincidence, it might also be possible that the rotational velocity of the observed galaxies relative to the rotational velocity of the Milky Way can lead to slight changes in the brightness of the galaxies, and consequently to a different number of objects as observed from Earth \citep{shamir2017large,shamir2020asymmetry,sym15061190}. It can also have a subtle yet consistent impact on the redshift as observed from Earth \citep{shamir2024simple}.

As shown with several different telescopes \citep{sym15061190}, galaxies that rotate in the same direction relative to the Milky Way are slightly dimmer than galaxies that rotate in the opposite direction relative to the Milky Way. Slight brightness differences between galaxies with opposite spin direction is expected due to the Doppler shift effect \citep{sym15061190}, but the difference is expected to be negligible. But it should also be remembered that the physics of galaxy rotation is one of the most tantalisingly complex phenomena, and its nature is not yet fully understood. If the brightness difference is significant, as shown empirically with data from SDSS \citep{sym15061190}, Pan-STARRS \citep{shamir2017large}, and HST \citep{shamir2020asymmetry}, it can lead to a difference between the number of galaxies that rotate with the same direction relative to the Milky Way, and in the opposite direction relative to the Milky Way as observed from Earth. That can lead to an asymmetry between the number of galaxies that rotate in opposite directions, and a dipole axis that peaks at around the Galactic pole. That explanation requires a modification in the physics of galaxy rotation, but as mentioned above, that physics is not yet fully known, and in fact is one of the most complex phenomena in nature. As discussed in \cite{shamir2024simple}, such asymmetry can also be related to other observed anomalies of brightness of objects, such as the unexpected cosmological-scale dipole anisotropy in the brightness of Ia supernovae \citep{PhysRevD.108.123533,cowell2023potential}.


The unprecedented imaging power of JWST provides a completely new look at the early Universe. The analysis shown here provides evidence that the number of galaxies spinning clockwise is significantly higher than the number of galaxies spinning counterclockwise. These observations are also aligned with previous observations using space-based and Earth-based instruments. Earth-based instruments also show evidence that in the opposite hemisphere the asymmetry is inverse, and form a dipole axis. A proposed experiment that would complement this study is the analysis of the corresponding field imaged by JWST in the opposite hemisphere. In that field a higher number of galaxies that rotate counterclockwise can expected. If an axis formed by the distribution of spiral galaxies exist, it might not be centred at Earth \citep{shamir2022new}, and therefore the distribution of spin directions in that field might not be exactly inverse to the distribution shown here. But if a higher number of counterclockwise galaxies is observed in that field, it would provide an indication of a consistent cosmological orientation towards a preferred direction, possibly forming a cosmological-scale axis. The proximity to the Galactic pole as well as to the CMB Cold Spot might also be possible directions for future research. JWST deep field images centred at the Galactic pole and at the CMB Cold Spot might provide additional information about the distribution of spiral galaxies in these parts of the sky to better understand the reasons leading to the anomaly.

\section*{Data Availability}

The main JWST deep field image used in the study is available at \url{https://webbtelescope.org/contents/media/images/01GXE4A07MB2RG6GHDGF3CHHJ4/}. SDSS galaxies analysed to reproduce \citep{hayes2017nature} are available at \url{https://people.cs.ksu.edu/~lshamir/data/sparcfire/}. Data used in \citep{psac058} are available at \url{https://people.cs.ksu.edu/~lshamir/data/iye_et_al}. Data used for comparing the magnitude difference of SDSS galaxies in the field around the galactic pole is available at \url{https://people.cs.ksu.edu/~lshamir/data/sdss_phot}. Data of annotated HST galaxies discussed in the paper are available at \url{http://people.cs.ksu.edu/~lshamir/data/assym_COSMOS}.

\section*{Acknowledgments}

I would like to thank the knowledgeable reviewer and associate editor for the insightful comments. The study was supported in part by National Science Foundation grant 2148878.


\bibliographystyle{apalike}
\bibliography{main_long}

\end{document}